\documentclass{IEEEtran}
\usepackage{amsmath}
\usepackage{amssymb}
\usepackage{graphicx}
\usepackage{tabularx}
\usepackage{bm}
\usepackage{cite}
\graphicspath{{./Figures/}} 
\DeclareGraphicsExtensions{.pdf,.jpg,.png,.jpeg,.eps}
\usepackage{enumitem}
\usepackage{fancyhdr}

\begin{document}
\title{How long is a resilience event in a transmission system?: Metrics and models driven by utility data}
\author{ Ian Dobson, Iowa State University \hspace{5mm} Svetlana Ekisheva, North American Electric Reliability
Corporation
\thanks{Ian Dobson is with ECpE dept., Iowa State University, Ames IA USA email: dobson@iastate.edu. Svetlana Ekisheva is with North American Electric Reliability
Corporation, Atlanta GA USA email: Svetlana.Ekisheva@nerc.net.
ID gratefully acknowledges support from USA NSF grant 2153163.}
}

\fancyhead[c]{\textnormal{ preprint to appear in IEEE Transactions on Power Systems, accepted July 2023 DOI: 10.1109/TPWRS.2023.3292328}}
\renewcommand{\headrulewidth}{0.0pt}
\fancyfoot[c]{\textnormal{\small This work is licensed under a Creative Commons Attribution 4.0 License; see http://creativecommons.org/licenses/by/4.0/}}

\maketitle	
\thispagestyle{fancy}

\begin{abstract}
\looseness=-1
We discuss  ways to measure duration in  a power transmission system  resilience event by modeling outage and restore processes from utility data.
We introduce novel Poisson process models that describe how resilience events progress and verify that they are typical using extensive outage data collected across North America.
Some usual duration metrics
show impractically high statistical variability, and we  recommend new duration metrics that perform better. Moreover, the Poisson process models have parameters that can be estimated from observed network data 
under different weather conditions, and are promising new models of typical resilience events.

\end{abstract}

\pagestyle{plain}
 
\section{Introduction}

Much of the analysis of electric power system resilience relies on describing the duration and magnitude of resilience events with quantitative metrics \cite{StankovicPS22,UmunnakweRSER21,PanteliProcIEEE17,NanRESS17,PoudelSJ19,ReedSYS09,WeiAM16,KellyGorhamEPSR20,CarringtonPS21}.  
The resilience events correspond to conditions of unusually high stress such as extreme weather or cascading and
are either simulated \cite{PanteliProcIEEE17,NanRESS17,PoudelSJ19} or extracted from historical data
\cite{ReedSYS09,WeiAM16,KellyGorhamEPSR20,CarringtonPS21}.
The metrics of duration and extent describe the performance of the power system as it responds to the high stress and, sometimes indirectly, the impact of the event on our society.
The metrics are broadly useful in improving the engineering of power system resilience, as evidenced by all the engineering references of this paper.
This paper addresses electric power transmission system metrics for the duration of resilience events and the durations of the outage and restore processes occurring within resilience events.
Here ``outage" refers to a  component being removed from service, and ``restore" refers to re-energizing a component to return it to service.

The duration of a resilience event would appear to be straightforward: 
The event starts with the first transmission outage at time $o_1$ and the event ends with the last restore  at time $r_n$, so that the event duration metric is simply $D_E=r_n-o_1$.
However, we will show that the timing $r_n$ of the last restore is so highly statistically variable that it is not meaningfully representative of the power system restoration.
(A metric is highly statistically variable if it is likely that its value can be much different than its estimated value, and we quantify this by the size of a confidence interval containing the estimate.)
Moreover, given the redundancy that is designed into power transmission systems, the last restore may have little or no impact on the power flowing to the distribution system and then to the customers.
Therefore we analyze a variety of  duration metrics to find new metrics which are less variable and more representative.

Our main approach is to develop new Poisson process models for the  outage and restore processes. The new models are driven by seven years of automatic outage data collected across North America by the North American Reliability Corporation (NERC) in its  Transmission Availability Data System (TADS).  
These statistical models enable the variability of the metrics to be quantified.
Moreover, parameters of the new models are closely related to some of the duration metrics.

This paper addresses the durations associated with transmission system resilience events in which there are substantial outages of transmission system elements. 
In particular, the paper does not address resilience events in which there are no outages or minimal outages, such as an extended heat wave that significantly limits transmission flows but causes no outages.
More generally, the paper is driven by outage and restore data for transmission system elements, and therefore does not address outages of generation, distribution system elements, and loads.

\subsection{Literature review}

Much of the previous work on statistical models of power system resilience events  addresses distribution systems.  Zapata \cite{ZapataIEEETandDCE08} models distribution system reliability with outages as a power-law Poisson process arriving at a  queue that is serviced by a power-law repair process to produce  a restore process. 
Wei and Ji \cite{WeiAM16} analyze distribution system resilience to particular severe hurricanes with a Poisson outage process  arriving at a  queue that repairs the outages to produce a restore process. 
Both the outage process rate and the repair time distribution vary in time 
as the hurricane progresses. 
Carrington \cite{CarringtonPS21} shows how to extract outage and restore processes from standard  distribution utility data.

Both \cite{WeiAM16} and \cite{ZapataIEEETandDCE08} statistically model the outage process and the component repair process,  and then calculate the restore process with a first-in-first-out queue model, whereas we  follow the insight of \cite{CarringtonPS21} in extracting and directly modeling the outage and restore processes.
Modeling the restore process directly from the data avoids the  complexities in queuing models of explicitly modeling the component repair and assuming an order of component repair.
While \cite{CarringtonPS21} fits the mean and standard deviation of the distribution system outage and restore processes to give a gamma distribution of restore times, it  does not give statistical process models as we do in this paper.
Moreover, the forms of the outage and restore processes  are quite different: for transmission systems the restore process dramatically slows over time and typically extends well beyond the end of the outage process, whereas in distribution systems the outage and restore processes overlap during most of the  event \cite{CarringtonPS21,WeiAM16}.

 Previous work also estimates individual component repair times  from distribution utility data.
For example, Jaech \cite{JaechPS19} predicts a gamma distribution of  individual component outage restoration times and customer hours lost with a  neural network, and
Liu \cite{LiuRESS08} fits generalized additive accelerated failure time models to hurricane and ice storm data. 

There is a continuing and very useful tradition of reliability analysis of bulk transmission systems 
that directly analyzes the average annual reliability of classes of components from observed data \cite{AdlerPD94,ChenPS05,BarkakatiPESGM19,EkishevaPD21,ZhouPS21}, or
applies steady state Markov analysis to calculate the average reliability \cite{BillintonAllanBook}.
The steady state Markov analysis has a huge literature with ingenious formulations to encompass different types of outage dependencies.
The contribution of extreme weather to the average reliability is modeled by having sets of Markov states 
corresponding to the extreme weather \cite{BillintonGTD06,BillintonJRR06}.
 \cite{JiangNAPS21} calculates the average steady state occurrence of compound outages from detailed outage data.
The present paper is different than steady state reliability approaches in that it analyzes transient systems-level processes of outage and restore rather than tracking individual components, and analyzes events of various sizes rather than reliability averaged over a year.

Most research on transmission system resilience uses conceptual frameworks and simulates physics-based models \cite{StankovicPS22,PanteliProcIEEE17,PanteliPS17,CiapessoniSG16,NanRESS17,KellyGorhamEPSR20}.
With the exception of  \cite{KellyGorhamEPSR20}, in which the simulation samples from empirically obtained distributions,
these approaches are not directly driven by observed data as in the present paper.

Cascading outages of transmission systems, which on an annual time scale\footnote{On a longer time scale,  complex system feedbacks produce a   ``statistical steady state" with the observed power law distribution of blackout size \cite{DobsonCH07,CarrerasPS16}.}
are rarer transient events involving a series of dependent outages, can be studied by extracting cascading events from outage data. The outage dependencies are diverse and 
can be caused by a common environment such as extreme weather 
as well as by interactions within the transmission system.
The statistics of cascading events can be studied by first extracting from observed outage data  the events in which outages bunch up and overlap.  For example, \cite{DobsonPS12,DobsonPS16} extracted events in the Northwest USA in which outages occurred in quick succession, and modeled the propagation and number of outages using a branching process, and how event outages spread in the network. 
\cite{MorrisPMAPS16} extracted events in Britain in which outages occurred in quick succession and analyzed their sizes and causes.
\cite{PapicAS20} extracted events from North American data with a quick succession of overlapping outages and analyzed their sizes and causes.
The present  paper uses the further refined event processing developed for processing and analyzing North American data in \cite{EkishevaPESGM21,EkishevaPMAPS22} and applied in the NERC state of reliability reports \cite{SOR21,SOR22}.

\subsection{Summary of paper contributions}

\noindent
This paper:
\begin{enumerate}
\item proposes new statistical models of outage and restore processes in transmission systems, and shows that the new models describe typical North American data.
\item analyzes statistical variability and interpretation of a variety of duration metrics.
\item recommends novel and more useful duration metrics.
\looseness=-2
\item reports typical values for model parameters and duration metrics for North America transmission  resilience events.
\end{enumerate}
 The previous conference papers and NERC reports \cite{EkishevaPESGM21,EkishevaPMAPS22,SOR21,SOR22} extract resilience events  from transmission system outage data and report the two duration metrics $D_{95\%}^{\ge}$ and $D_n$ for the larger or largest events. 
 The fruitful previous applications of these duration metrics motivate in this paper the extensive new analysis of a range of duration metrics and the recommendations backed by this analysis of better performing duration metrics.
The extraction of the transmission system resilience events developed in \cite{EkishevaPESGM21,EkishevaPMAPS22} is not the subject of this paper, but since it is used in the data processing of this paper, we specify in section II the precise version of the event extraction used.
Section II also summarizes the outage data used in the paper and states and briefly comments on the definitions of the outage  and restore processes \cite{CarringtonPS21} since this paper uses these processes.

 The duration of resilience events has clear importance to the public, engineers, regulators, and policy makers.
This motivates our consideration of the performance of a range of duration metrics.
We are not aware of another paper addressing the question of how duration metrics perform, and we approach the question with novel methods.
In particular,
the stochastic models 
of typical transmission system resilience processes proposed and validated with 
extensive data in the paper are novel, and we expect that these new models will be useful well beyond 
this paper's more immediate goal of proposing and analyzing better duration metrics.

\section{Resilience events and processes}
\label{processes}
To obtain resilience metrics from utility outage data, we first need to automatically extract resilience events and the outage and restore processes for each event. 
This section explains how to do this based on previous work \cite{CarringtonPS21,EkishevaPESGM21,EkishevaPMAPS22} and establishes the notation needed for the paper.

\subsection{Utility data and extracting resilience events}
NERC’s TADS collects outage and inventory data for the following four types of transmission elements: AC circuits, transformers, AC/DC back-to-back converters, and DC circuits \cite{TADSDRI} which are part of the North American Bulk Power System (i.e. operated at 100 kV or higher) \cite{NERCBES18}.
The detailed automatic outage data include the outage and restore time to the nearest minute, the initiating cause code for each outage, and the sustaining cause code for sustained outages. In this paper we analyze the approximately 62 000 automatic outages for all elements reported in TADS from 2015 to 2021 for the Eastern, Western, and ERCOT interconnections.

 \looseness=-1
 A key step in resilience analysis of real data is automatically extracting resilience events.
 For each interconnection, the  automatic outages  are grouped together into resilience events based on the bunching and overlaps of their starting times and
 durations. We quote from \cite{EkishevaPMAPS22} the algorithm used: ``Every outage in an event has to either start within five minutes of a previous outage in the event or overlap in duration with at least one previous outage in the event that has a difference in starting time not exceeding one hour. In applying this algorithm, repeated momentary outages of the same element are neglected if they occur within 5 minutes of each other."
 We use this algorithm to automatically group outages into resilience events (their sizes vary from 1 to 352 outages) and then analyze all the resilience events with 10 or more outages.
 An event that contains at least one outage with a weather-related initiating or sustained cause code is defined as a weather-related event. The weather-related TADS cause codes are lightning, weather excluding lightning, fire, and environmental. This procedure identified 352 transmission events with 10 or more outages, 329 of which are weather-related. Note that events are defined so that if an outage is included in an event, then so is its corresponding restore. Therefore the number of outages in an event is equal to the number of restores.

\subsection{Outage, restore, and performance  processes}
Suppose that  the resilience event has $n$ outages  at times $o_1\le o_2\le...\le o_n$ and $n$ restores at times $r_1\le r_2\le...\le r_n$.
Note that the outages are sorted into the order in which the outages occur, and the 
 restore times are sorted into the order in which the restores occur. This sorting implies  that the $k$th restore time $r_k$ is not usually the restore of the $k$th outage $o_k$.

\begin{figure}[htpb]
	\centering
	\includegraphics[width=\columnwidth]{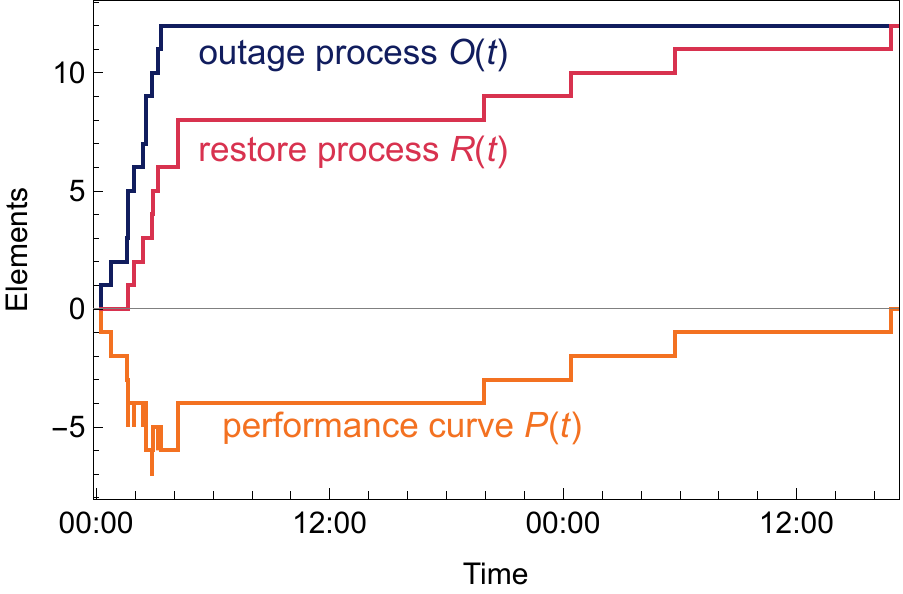}
	\caption{Processes  for a transmission system resilience event with 12 outages.}
	\label{fig:resiliencecurveplot}
\end{figure}

For each event, the outage process $O(t)$ is the cumulative number of outages at time $t$ and the restore process $R(t)$ is the cumulative number of outages at time $t$:
\begin{align}
O(t)&=\mbox{ number of outages $o_j$ with }o_j\le t
\label{cumulativeoutages}\\
R(t)&=\mbox{ number of restores $r_k$ with }r_k\le t
\label{cumulativerestores}
\end{align}
Both processes start at zero at the beginning of the event and increase to the total number of outages $n$, as can be seen in the example in Fig.~\ref{fig:resiliencecurveplot}.

\looseness=-1
Resilience studies \cite{StankovicPS22,PanteliProcIEEE17,NanRESS17,PoudelSJ19} often define for each event a performance (or resilience) curve $P(t)$, which is the negative of the cumulative number of {\sl unrestored} outages at time $t$. The performance curve decrements for each outage and increments for each restore as shown in Fig.~\ref{fig:resiliencecurveplot}.
Indeed, the performance curve is related to the outage and restore processes by 
$    P(t)=R(t)-O(t)$.
The performance curve can be uniquely decomposed into its outage and restore processes, and  it contains the same information as the outage and restore processes \cite{CarringtonPS21}.

The outage and restore processes, while straightforward, are fundamental to analyzing real outage data, and they have several distinctive features \cite{CarringtonPS21}:
(a) The outage and restore processes routinely overlap in time in real data; this differs from the customary idealized outage and restore phases of resilience that are separated in time \cite{StankovicPS22,PanteliProcIEEE17,NanRESS17,PoudelSJ19,KellyGorhamEPSR20}. 
(b) The analysis is at a systems level and is not focused on tracking individual elements: it only counts the numbers of outages and restores and it does not track which outaged element restored when or the order in which  elements restore.
(c) The forms of the outage and restore processes and performance curve readily lead to resilience metrics that describe each process; in particular, it is useful to have separate metrics describing the outage process and the restore process.

\begin{figure*}[t]
	\includegraphics[width=7.17in]{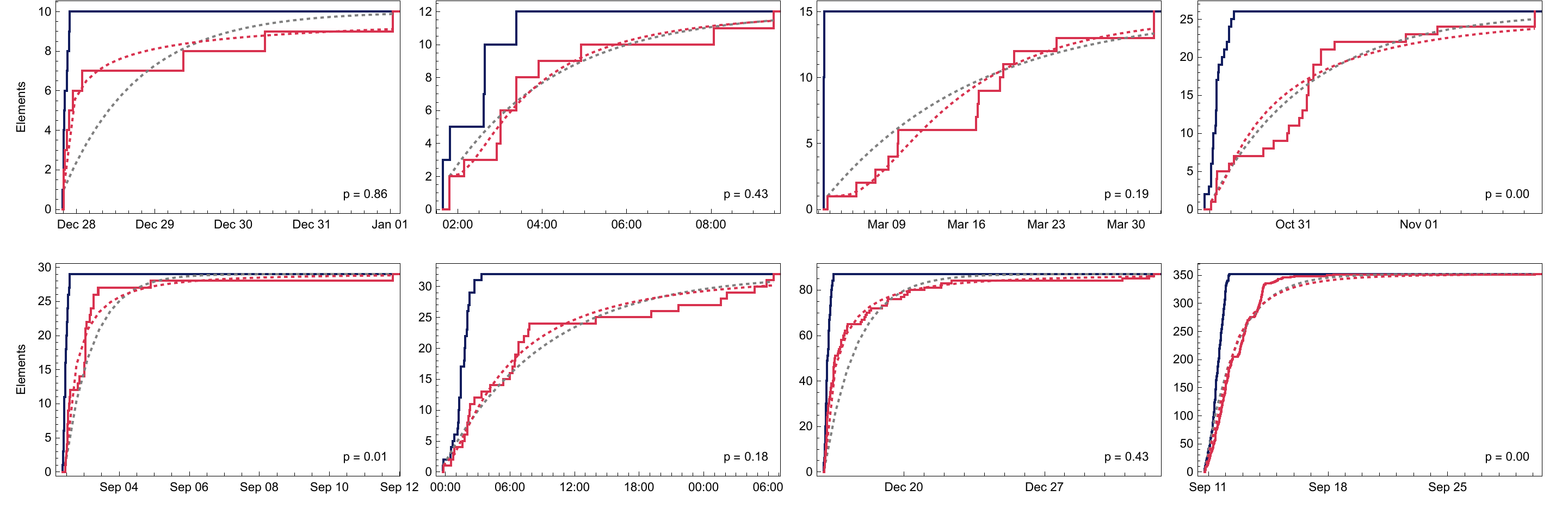}
	\caption{Examples of outage processes (dark blue) and restore processes (red) for events. Red dashed line is lognormal restore approximation, gray dashed line is exponential restore approximation. p-value is from Anderson Darling test on lognormal fit to restore process.}
	\label{lognormalfits}
	\vspace{-5mm}~
\end{figure*}

\begin{figure}[ht]
	\centering
	\includegraphics[width=\columnwidth]{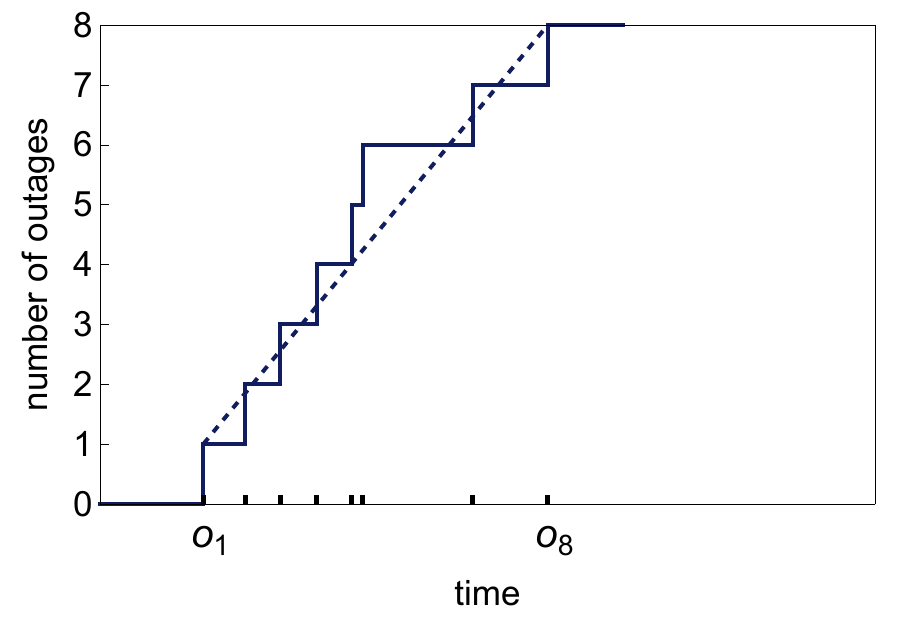}
	\caption{Horizontal axes ticks show eight outage times $o_1$,$o_2$,...,$o_8$ produced by a Poisson process with constant rate $\lambda_O$.
	The resulting outage process $O(t)$ is the dark blue stepped line. $O(t)$ is approximated by the average outage process $\overline{O}(t)$, which is the dashed line of slope  $\lambda_O$.}
	\label{fig:linearPoisson}
\end{figure}
\begin{figure}[ht]
	\centering
	\includegraphics[width=\columnwidth]{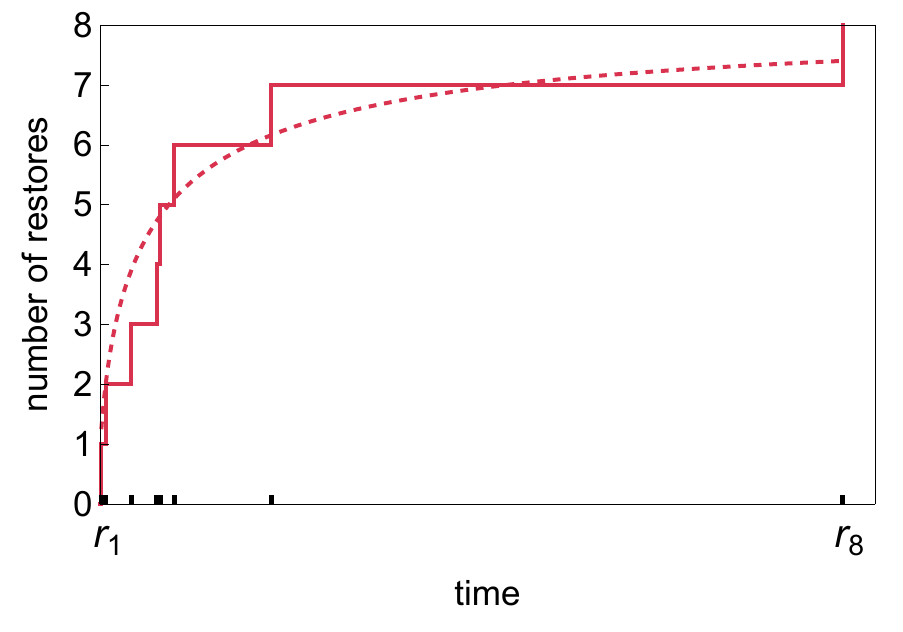}
	\caption{\looseness=-1 Horizontal axes ticks show eight restore times $r_1$,$r_2$,...,$r_8$ produced by a Poisson process with  lognormal rate. The resulting restore process $R(t)$ is the red stepped line. $R(t)$ is approximated by the average restore process $\overline{R}(t)$, which is the dashed curve.  $\overline{R}(t)$ is proportional to  the CDF of the lognormal distribution and its slope is the Poisson process rate. }
	\label{fig:lognormalPoisson}
\end{figure}

\section{Poisson process models of outage and restore}

\looseness=-1
This section introduces new Poisson process models that describe typical outage and restore processes in our transmission system data. Fig.~\ref{lognormalfits} shows examples.
The mean values of these Poisson processes are a useful approximation of the outage and restore processes.
Moreover, parameters of the Poisson process models yield  resilience metrics, and section \ref{vary} uses the Poisson process models to quantify the variability of the metrics.
We consider two different Poisson models for the restore process, based on lognormal and exponential rates respectively.
The fit of the Poisson models with the data is discussed in section \ref{fit}, where it is shown that the model with a lognormal rate typically fits the restore process  better than the model with an exponential rate.

\subsection{Poisson process of outage times with constant rate}
\label{Poissonoutage}

The data for each event specifies that there are $n$ outages in the event and that the outages start at time $o_1$ and end at time $o_n$. Given this information, and assuming a constant rate Poisson process,
we model the outage times as occurring randomly and at a constant rate $\lambda_O$ in the time interval $(o_1,o_n)$. 
In particular, given that there are $n$ outages in $(o_1,o_n)$, the $n-2$ outage times
$o_2,...,o_{n-1}$ are independent samples from a uniform distribution on $(o_1,o_n)$ sorted into ascending order\footnote{One well known property of a constant rate Poisson process  is that, if there are a given number of  outages in an interval, then these outage times are uniformly distributed in that interval \cite[Thm.~4A, Ex. 4A]{Parzen15},
\cite[Thm.~5.2]{Ross07}.}.

A metric characterizing the outages is their rate $\lambda_O$, which is estimated for each event as\footnote{
Since there are $n-1$ time differences between the $n$ outages, the estimated average time difference between successive outages is 
$(o_n-o_1)/(n-1)$, and then the estimated rate $\lambda_O$ is the reciprocal of the average time difference.}
\begin{align}
    \lambda_O=\frac{n-1}{o_n-o_1}
   \label{lambdaOhat}
\end{align}
The average or expected cumulative number of outages $\overline{O}(t)$ at time $t$ is 
\begin{align}
   \overline{O}(t)={\rm E}[O(t)]=1+\lambda_O (t-o_1), \qquad o_1\le t\le o_n
   \label{averageO}
\end{align}
$\overline{O}(t)$ approximates the outage process $O(t)$ as shown in Fig.~\ref{fig:linearPoisson}.
We see in Fig.~\ref{lognormalfits} some typical examples in which the cumulative number of outage increases in the  linear way given by (\ref{averageO}).
The total number of outages is $\overline{O}(o_n)=n$.
For each event, $\lambda_O$ can be estimated from (\ref{lambdaOhat}), and then
the averaged outage process (\ref{averageO}) approximates and describes the outage process $O(t)$.

\subsection{Poisson process of restore times with lognormal rate}
\label{Poissonlognormal}

The data for each event specifies that there are $n$ restores in the event and that the restores start at time $r_1$. 
We work with the restore times relative to $r_1$; that is, $r_j-r_1$, $j=1,2,..,n$.
The first restore time relative to $r_1$, and any other simultaneous restores at $r_1$, become $r_1-r_1=0$.
Suppose that first restore that occurs at a time $>r_1$ is $r_{z+1}$. 
Usually $r_2>r_1$ and $z=1$. 

 The restore times typically happen with a rate that varies,
as can be seen in the 
examples in Fig.~\ref{lognormalfits}.
In particular, the rate of restores typically slows dramatically for the final restores.
We model the $n\!-\!z$ positive restore times $r_j\!-\!r_1$, $j={z+1}, {z+2},...,n$ as occurring randomly in a nonhomogeneous Poisson process at a rate proportional to a lognormal distribution.
In particular, given that there are $n-z$ outages in the time interval $(r_1,\infty)=\{t~|~t>r_1\}$, the $n-z$ restore times
$r_{z+1}\!-\!r_1,...,r_n\!-\!r_1$ are independent samples from a lognormal distribution on  $(r_1,\infty)$ sorted into ascending order. 
There are some extremely long restore times $r_n$ in the data (up to a year is recorded), and this is reflected in the modeling of the process as unbounded in $(r_1,\infty)$.

Let the lognormal distribution have parameters $\mu$ and $\sigma$ and probability density function $f_{\mu,\sigma}(t)$.
Then the Poisson process rate is proportional to the probability density function:
\begin{align}
    \lambda_R(t)=
    (n-z) f_{\mu,\sigma}(t-r_1),\quad t> r_1
    \label{lambdaR}
\end{align}

By definition of the lognormal distribution,
since the restore times
$r_{z+1}-r_1,r_{z+2}-r_1,...,r_n-r_1$ are independent samples from a lognormal distribution,  the natural logarithms of the restore times $\ln(r_{z+1}-r_1),\ln(r_{z+2}-r_1),...,\ln(r_n-r_1)$ are independent samples from a normal distribution.
The standard parameters characterizing the lognormal distribution  are the mean $\mu$ and standard deviation $\sigma$ of the normal distribution. Therefore we estimate $\mu$ and  $\sigma$ for each event by 
\begin{align}
\mu&=\frac{1}{n-z}\sum_{k=z+1}^n \ln(r_k-r_1)\label{muhat}\\
\sigma^2&=\frac{1}{n-z-1}\sum_{k=z+1}^n (\ln(r_k-r_1)-\mu)^2
\label{sigmahat}
\end{align}

The Poisson process restore rate $\lambda_R(t)$ is proportional to the lognormal distribution  as shown in (\ref{lambdaR}).
Then the average or expected cumulative number of restores $\overline{R}(t)$  is 
\begin{align}
   \overline{R}(t)
   &={\rm E}[R(t)]
   =z+\int_{r_1}^t\lambda_R(\tau)d\tau\nonumber\\
   &=z+(n-z)\int_{r_1}^t f_{\mu,\sigma}(\tau-r_1)d\tau\label{prop}\\
   &=z+(n-z)\frac{1}{\sigma\sqrt{2\pi}}\int_{-\infty}^{\ln (t-r_1)} e^{-\frac{(y-\mu)^2}{2\sigma^2}}dy\nonumber\\
 &=z+(n-z)\Phi\bigg[ \frac{\ln (t-r_1)-\mu}{\sigma}\bigg]
 ,\quad t\ge r_1
   \label{averageR}
\end{align}
where $\Phi$ is the CDF of the standard normal distribution.
Equation (\ref{prop}) shows that  $\overline{R}(t)-z$ is proportional to the CDF of the lognormal distribution, and (\ref{averageR}) expresses  $\overline{R}(t)$ in terms of the parameters $\mu$ and $\sigma$.
$\overline{R}(t)$ approximates the restore process $R(t)$ as shown in Fig.~\ref{fig:lognormalPoisson}.

\looseness=-1
The lognormal model has parameters $\mu$, $\sigma$,  $z$, and $n$.
For each event, $\mu$ and $\sigma$ can be estimated from (\ref{muhat}) and (\ref{sigmahat}) and then
the averaged outage process $\overline{R}(t)$  (\ref{averageR}) approximates and describes the restore process $R(t)$.
Examples of the approximating restore curves are shown by red dashed lines in Fig.~\ref{lognormalfits}.

\subsection{Poisson process of restore times with exponential rate}
\label{Poissonexp}

We can substitute the exponential distribution for the lognormal distribution of  subsection~\ref{Poissonlognormal} to obtain a Poisson restore process with exponential rate.
That is, given that there are $n-z$ outages in $(r_1,\infty)$, the $n-z$ restore times
$r_{z+1}\!-\!r_1,...,r_n\!-\!r_1$ are independent samples from an exponential distribution on  $(r_1,\infty)$ sorted into ascending order.
We analyze the exponential restore rate because it is an analytically convenient choice to try to describe the slowing rate of restores.

Let the exponential distribution have time constant $\tau$ and probability density function $
\tau^{-1}\, e^{- t/\tau}$ for $t\ge0$.
Then the Poisson process rate is 
\begin{align}
    \lambda_{R\rm exp}(t)=
    (n-z) \tau^{-1}\, e^{- (t-r_1)/\tau}, \quad t> r_1
    \label{lambdaRexp}
\end{align}
and the expected cumulative number of restores  is 
\begin{align}
   \overline{R}_{\rm exp}(t)
   &=z+(n-z)\int_{r_1}^t  \tau^{-1}\, e^{- (s-r_1)/\tau}ds\\
   &=z+(n-z)[1-e^{- (t-r_1)/\tau}],\quad t\ge r_1
   \label{averageRexp}
\end{align}

We estimate the exponential time constant by 
\begin{align}
    \tau=\frac{1}{n-z}\sum_{k=z+1}^n(r_k-r_1)
    \label{tauhat}
\end{align}
$\tau$ is the arithmetic mean of the positive restore times relative to $r_1$.
The exponential model has parameters $\tau$, $z$, and $n$.
For each event,  $\tau$ can be estimated from  (\ref{tauhat}), and then
the averaged outage process $\overline{R}_{\rm exp}$ in (\ref{averageRexp}) approximates and describes  the restore process $R(t)$.
Examples of the approximating restore curves are shown by gray dashed lines in Fig.~\ref{lognormalfits}.

\section{Duration metrics}
\label{duration}

There are many possible metrics describing durations in resilience events. 
 This section defines and describes a variety of these metrics.

\subsection{Straightforward duration metrics}

\begin{description}
\item[outage duration] $D_O=o_n-o_1$
\item[time to first restore] $D_{r1}=r_1-o_1$
\item[restore duration] $D_n=r_n-r_1$
\item[restore time to $k$th  restore] $D_k=r_k-r_1$
\item[event duration]  $D_E=r_n-o_1$
\end{description}

The outage process starts at the first outage $o_1$ and ends at $o_n$ so that the outage duration $D_O=o_n-o_1$. The first restore is at time $r_1$ and 
the time to the first restore is $D_{r1}=r_1-o_1$. That is, $D_{r1}$ quantifies how much the start of the restore process is delayed.
The restore process starts at $r_1$ and ends at the last restore $r_n$ so that the restore duration $D_n=r_n-r_1$.
The event starts at time $o_1$ and ends at time $r_n$.
The event duration $D_E=r_n-o_1$ can be split into the time to the first restore  and the restore duration:
\begin{align}
D_E=r_n-o_1=(r_1-o_1)+(r_n-r_1)=D_{r1}+D_n
\label{DE}
\end{align}

This section discusses restore duration, but
 the corresponding  metrics describing  event duration are easily obtained from the metrics for restore duration by adding the time to first restore $D_{r1}$ as in (\ref{DE}).
The outage duration $D_O$ and time to first restore $D_{r1}$ are useful metrics, but section \ref{discuss} explains that the restore duration $D_n$ and the event duration $D_E$ suffer from high variability.

\subsection{Restore metrics based on quantiles}
It is of interest to quantify the time to reach a given percentage $x$ of restoration, or, equivalently, the $x/100$ quantile of the restore times  $0,r_2-r_1,r_3-r_1,...,r_n-r_1$.
There are many different definitions of quantiles (\cite{HyndmanAS96} analyzes 10 definitions used in statistics), and correspondingly many ways to define restore metrics based on quantiles.
This subsection discusses two metrics of restore duration based on quantiles; the first metric quantizes to a  restore time while the second metric interpolates between restore times.

\vspace{1mm}
\noindent
{\bf time to first restore with at least $x\%$ restoration}
\begin{align}
    D_{x\%}^{\scriptscriptstyle\ge}=
    r_{\lceil nx/100\rceil}-r_1
\end{align}
The ceiling function ${\lceil u\rceil}$ is the smallest integer $\ge u$. For example, $D_{95\%}^{\scriptscriptstyle\ge}$ is the time between the first restore $r_1$ and the first restore $r_{\lceil 0.95n\rceil}$ at which at least 95\% of the restores are completed.
It follows that $D_{95\%}^{\scriptscriptstyle\ge}=D_n$ for $n<20$,  $D_{95\%}^{\scriptscriptstyle\ge}=D_{n-1}$ for $20\le n<40$, and $D_{95\%}^{\scriptscriptstyle\ge}=D_{n-2}$ for $40\le n<60$. For example, for $n=16$, $\lceil 0.95 n\rceil=\lceil 15.2\rceil=16$ and $D^{\ge}_{95\%}=D_{16}$. These quantum jumps in $D_{95\%}^{\scriptscriptstyle\ge}$ as  $n$ varies, and which also occur as $x$ varies, are unsatisfactory when analyzing a range of events. This can be fixed with the following  more elaborate quantile definition.

\vspace{1mm}
\noindent
{\bf 
restore time to $x\%$ of restoration}
\begin{align}
    D_{x\%}&=
   (1- (u-\lfloor u\rfloor))r_{\lfloor u\rfloor}+(u-\lfloor u\rfloor)r_{\lceil u\rceil}-r_1
   \notag
    \\
    &=
   (1- (u-\lfloor u\rfloor))D_{\lfloor u\rfloor}+(u-\lfloor u\rfloor)D_{\lceil u\rceil}
    \label{quantiledefn}\\
\mbox{where~~}&
u=\min\Big\{\frac{1}{3}+\big(n+\frac{1}{3}\big)\frac{x}{100},\,n\,\Big\}
\label{u}
\end{align} 
The ceiling function $\lceil u\rceil$ is the smallest integer $\ge u$, the floor function ${\lfloor u\rfloor}$ is the largest integer $\le u$, and $u-\lfloor u\rfloor$ is the fractional part of $u$. 

Eqn.~(\ref{quantiledefn}) shows that $D_{x\%}$ linearly interpolates  between restore times   $D_{\lfloor u\rfloor}$ and $D_{\lceil u\rceil}$.
$D_{x\%}$ uses the median-based quantile 
definition\footnote{implemented in R as quantile type 8, and in Mathematica by Quantile with parameters $\{\{1/3, 1/3\}, \{0, 1\}\}$} recommended by  \cite{HyndmanAS96}, but also limits $u$ to a maximum of $n$ in (\ref{u}). When  limiting applies, $D_{x\%}=D_n$. 

In contrast to $D^{\ge}_{x\%}$,
$D_{x\%}$ changes continuously as $x$ varies and with much smaller jumps as $n$ varies. For this reason, we strongly prefer $D_{x\%}$ to $D^{\ge}_{x\%}$.

$D_{50\%}$ evaluated with (\ref{quantiledefn})  reduces to the usual median. That is,
letting $\ell=\lceil n/2 \rceil$,
\begin{align}
\hspace{-0.2in}
    D_{50\%}=
    &\begin{cases}
    r_{\ell}-r_1 \hspace{0.1in}  &,n=2\ell-1= \mbox{odd}\\
    \frac{1}{2}(r_{\ell}+r_{\ell+1})-r_1 &,n=2\ell=\mbox{even}\\
    \end{cases}
    \label{mediandefn}
\end{align}

\subsection{Metrics related to restore process models}
\label{restoremetrics}
These metrics work with the positive restore times relative to $r_1$; that is, $r_j-r_1$, $j=z+1,z+2,..,n$.\footnote{The following metric definitions require a positive outage duration ($o_n>o_1$) so that $z<n$. 
 If $o_n=o_1$, we define the metric to be zero.} 
Usually $z=1$ as explained in section \ref{Poissonlognormal}.
\begin{description}
\item[geometric mean of positive restore times] ~\\$D_{\rm GM}=\left[\prod_{k=z+1}^n(r_k-r_1)\right]^\frac{1}{n-z}=e^{\mu}$
\item[arithmetic mean of log restore times]
~\\$\mu=\frac{1}{n-z}\left[\sum_{k=z+1}^n\ln[r_k-r_1]\right]=\ln D_{\rm GM}$
\item[standard deviation of log restore times]
~\\$\sigma=\sqrt{\frac{1}{n-z-1}\sum_{k=z+1}^n(\ln[r_k-r_1]-\mu)^2}$
\item[restore time to $x\%$ restoration assuming lognormal]~\\
$ D_{x\%}^{\rm ln}$ satisfies $nx/100= \overline{R}(D_{x\%}^{\rm ln}+r_1)$ and\\
$
 n x/100-z
=(n-z)\Phi[ (\ln D_{x\%}^{\rm ln}-\mu)/\sigma]
$
so that 
\begin{align}
 D_{x\%}^{\rm ln}={\rm exp}\big[\mu+ \sigma\,\Phi^{-1}\Big(\frac{nx/100-z}{n-z}\Big)\big]
\label{Dxpercent}
\end{align}
Note that $ D_{(50+50z/n)\%}^{\rm ln}=e^{\mu}=D_{\rm GM}$.

\item[arithmetic mean of nonzero restore times]~\\
$  \tau=\frac{1}{n-z}\sum_{k=z+1}^n(r_k-r_1)$

\item[restore time  to $x\%$  restoration assuming exponential]~\\
$D_{x\%}^{\rm exp}$
 satisfies
$nx/100-z= \overline{R}_{\rm exp}(D_{x\%}^{\rm exp}+r_1)$ 
and~\\
$
 n(1-x/100)
=(n-z)\exp[-D_{x\%}^{\rm exp}/\tau]
$ so that 
\begin{align}
D_{x\%}^{\rm exp}=\tau\ln \Big[\frac{n-z}{n(1-x/100)}\Big]\notag
\end{align}
\end{description}
\noindent
The  average restoring half life $ D_{50\%}^{\rm exp}=\tau\ln [2(\frac{n-z}{n})]$ is the average time 
for the number of unrestored outages to halve averaged over the restore process assuming exponential decay.

 There are variants of $D_{x\%}^{\rm ln}$ and $D_{x\%}^{\rm exp}$ with  slightly simpler formulas that describe the time to  restoration of $x\%$ of the $n\!-\!z$ nonzero restore times. 
 For these variants, $D_{x\%}^{\rm ln}$ becomes ${\rm exp}[\mu+ \sigma\,\Phi^{-1}(x/100)]$ and $D_{x\%}^{\rm exp}$ becomes $\tau\ln [1/(1-x/100)]$. 
 We prefer the definitions of $D_{x\%}^{\rm ln}$ and $D_{x\%}^{\rm exp}$ above because the time to restoration of $x\%$ of all $n$ restore times seems more straightforward.
 
 All the duration metrics in the paper (labelled with $D$)  are given in hours so that the time unit $t_u=1$ hour.
 We now discuss the units of $\mu$ and $\sigma$. 
 A more precise version of 
 $\mu=\ln D_{\rm GM}$ is $\mu=\ln (D_{\rm GM}/t_u)$ (or
 $ D_{\rm GM}=t_u e^{\mu}$).
 Dividing $D_{\rm GM}$ in hours by $t_u=1$ in hours gives the required nondimensional argument of the logarithm \cite{MattaACS11}.
Changing $t_u$ will cause a change in the value of $\mu$. $\sigma$ does not depend on the units used and gives the same value for any choice of $t_u$. 

\section{Discussing restore metrics $D_n$, $D_{\rm GM}$, $D_{95\%}$, $D_{95\%}^{\rm ln}$}
\label{discuss}

\begin{table*}[htbp]
	\caption{Summary of metrics, recommendations, and typical values}
	\label{summary}
	\centering
	\begin{tabular}{ clclc }
		\multicolumn{2}{c}{metric~~~~~~~~~~~~} &recommend?&comment&median\\
		\hline
		$n$&number of outages/restores&Yes& useful measure of event size&13.5
		{\vrule height 9pt depth 0pt width 0pt}\\
		$D_O$&outage duration  & Yes& useful description of outage process&2.69	\\
		$\lambda_O$&outage rate  & Yes& useful description of outage process &5.45\\
		$D_{r1}$&time to first restore  & Yes& useful description of delay in start of restores&0.52 \\[1 mm]
		$D_{E}$&event duration& No&$=D_{r1}+D_{n}$; extreme variability&69.8\\
		$D_{n}$&restore  duration (time to last restore)& No&extreme variability&69.1\\
		$D_{n\!-\!1}$&restore time to $(n\!-\!1)$th restore&No&{$D_{95\%}$ preferred}&31.4\\[1 mm]
		$D_{95\%}^{\scriptscriptstyle \ge}$&first restore time with $\ge 95\%$ restore&No&$D_{95\%}$  preferred since continuous&55.4\\
		$D_{90\%}$&restore time to $90\%$ quantile & Yes&&39.2\\
		$D_{95\%}$&restore time to $95\%$ quantile & Yes&  $D_{95\%}^{\rm ln}$ is an alternative&65.2\\[3pt]
		$\mu$&mean of log restore times& 
				No&$\mu=\ln D_{\rm GM}$ and  $D_{\rm GM}$ is recommended.&1.64\\
		$\sigma$&standard deviation of log restore times&--&&1.56\\
		$D_{95\%}^{\rm ln}$&restore time to $95\%$ with lognormal$(\mu,\sigma)$&--&
	$D_{95\%}$ slightly preferred; lognormal fit only typical &67.7\\[1mm]
				$\tau$&arithmetic mean of  restores$>0$; exp time constant\hspace{-5mm} & No&exponential fit poorer; variable for small $n$ &16.4\\
		$D_{95\%}^{\rm exp}$&restore time to $95\%$ with exponential$(\tau)$&No&exponential fit poorer&47.8\\[1mm]
		$D_{50\%}$&median restore time&No&
		$D_{\rm GM}$ preferred&4.27\\
		$D_{\rm GM}$&geometric mean of restore times&Yes & best, least variable restore performance metric;&5.15\\[-1mm]
		&&& also estimates median of restores$>0$
		{\vrule height 8pt depth 5pt width 0pt}
		\\
					        		\hline	\\[-3mm]
			\multicolumn {5}{r}{all durations in hours, $\lambda_O$ in per hour}
	\end{tabular}
		\vspace{-3mm}
\end{table*}

All duration metrics of the restore process  are subject to substantial statistical variability that can undermine their usefulness, especially for smaller values of event size $n$. The variabilities of the restore  metrics are analyzed in section~\ref{vary} by calculating the size of their confidence interval, and only the conclusions about their variability are stated here.

The  restore duration metric
 $D_n$ is straightforward, but it is typically too highly variable to be a reliable estimate. Moreover, $D_n$ depends strongly on the last or last few restores, preventing $D_n$  from describing the performance throughout the entire restore process.
 This dependence also makes $D_n$ relate poorly to transmission performance  because these last restores may be unimportant for customers, or may be excessively delayed by factors out of the control of the utility, such as the difficulty of repairing transmission lines in the mountains in the winter or structural damage caused by hurricane or tornado.  

\looseness=-1
The geometric mean of the positive restore times 
$D_{\rm GM}$ is the best  estimate of restore performance in terms of having the least variability.
It is also clear that $D_{\rm GM}$ depends on all the restores throughout the restore process.
We now discuss how $D_{\rm GM}$ also estimates a median of the restore process.
Since the normal distribution is symmetrical about its mean value, the mean $\mu$ also estimates the median of the normal distribution, and therefore $D_{\rm GM}=e^\mu$ estimates the median of the lognormal distribution\footnote{Only the symmetry of the distribution of the logarithm of the nonzero restore times relative to $r_1$ is needed here.}.
In fact,  $D_{\rm GM}$ is a better estimate (less variance) of the median than applying the standard formula (\ref{mediandefn}) for the median. 
The detailed correspondence is that $D_{\rm GM}$ estimates the median of $r_j-r_1$, $j=z+1,z+2,...,n$, which is modestly greater than\footnote{
For $z=1$, difference in the medians is $(r_{\ell+1}-r_{\ell})/2$, where $\ell=\lceil n/2 \rceil$.} the median of all the restore times $r_j-r_1$, $j=1,2,..,n$ calculated in (\ref{mediandefn}).  That is, under the lognormal model, $D_{\rm GM}$ is a good estimate of the median of the positive restore times relative to $r_1$, and approximates from above the median $D_{50\%}$ of all restore times relative to $r_1$.

While $D_{\rm GM}$ is an informative metric with the lowest variability, $D_{95\%}$ and $D_{95\%}^{\rm ln}$ can be used as more representative of the almost complete duration of the restore process, with the compromise of higher variability than $D_{\rm GM}$.
$D_{95\%}$ is a more smoothly varying quantile metric indicating the 95\% completion of the restore process.
$D_{95\%}^{\rm ln}$ is also smoothly varying.
$D_{95\%}$ is a bit more variable than $D_{95\%}^{\rm ln}$, particularly for small $n$.
Overall, we slightly prefer
$D_{95\%}$ to $D_{95\%}^{\rm ln}$ because the quantile 
approach is less model dependent, whereas $D_{95\%}^{\rm ln}$ will work best in the typical lognormal restore case.

Table~\ref{summary} summarizes the metrics and our recommendations.

\section{Typical values of metrics \& model parameters}

Typical values of metrics and parameters are given for all the data in
Table~\ref{summary} and for each interconnection in
Table \ref{typical}; these values are expected to be useful for modeling and assessing interconnection-specific transmission events. 
Due to the heavy tails in their distribution, some quantities in Table \ref{typical} such as $D_n$ have mean values that greatly exceed the median and large standard deviations. 
In these cases, the estimated mean has substantial statistical variation and poorly indicates a typical value; the median is a better typical value.
The large standard deviations arise from both the metric statistical variability and the metric variation between events.

\begin{table}[hbpt]
	\caption{Typical values of metrics by interconnection}
	\label{typical}
	\centering
	\setlength{\tabcolsep}{0.3em}
	\begin{tabular}{ cccccccccc }
		&\multicolumn{3}{c}{Eastern}&\multicolumn{3}{c}{ERCOT}&\multicolumn{3}{c}{Western} \\
		Metric~~&mean&\!SD&\!\!median&~~mean&\!SD&\!\!median&~~mean&\!SD&\!\!median \\
		\hline
		$n$  & 23.2&38.2&13&16.9&10.0&13&20.1&17.7&14  \\
		$D_O$   & 3.5&3.6&2.8&2.6&2.1&2.3&2.8&2.3&2.5  \\
		$\lambda_O$   & 7.3&8.6&5.1&6.5&3.7&5.2&24.0&99.0&6.4  \\
		$D_{r1}$   & 0.78&1.07&0.53&1.28&1.34&0.95&0.65&0.80&0.43  \\
		$D_{E}$ & 379&1088&73&154&227&53&219&494&62 \\
		$D_{n}$ & 379&1088&72&153&228&50&218&494&61 \\
		$D_{n\!-\!1}$ &126&332&32&75&81&36&81&210&26\\
		$D_{95\%}^{\ge}$ & 305&1000&62&128&204&49&170&438&46\\
		$D_{90\%}$ & 151&471&44&78&83&39&103&262&32\\
		$D_{95\%}$ & 294&945&67&122&182&49&180&442&48\\
		$\mu$ & 1.68&1.17&1.76&1.48&1.62&2.19&1.23&1.11&1.10\\
		$\sigma$ & 1.64&0.57&1.59&1.56&0.59&1.67&1.57&0.65&1.46\\
		$D_{95\%}^{\rm ln}$&397&2740&77&199&327&55&132&362&46\\
		$\tau$&49.8&154&17.6&28.6&30.8&15.0&28.5&57.6&12.5\\
		$D_{95\%}^{\rm exp}$&145&449&52&84&90&44&83&167&37\\
		$D_{50\%}$ & 15.3&65.3&4.8&18.1&28.8&5.3&5.6&6.1&2.6\\
		$D_{\rm GM}$ &12.8&51.6&5.8&10.0&9.5&8.9&5.8&5.6&3.0\\
					        		\hline		\\[-3mm]
					        	&	\multicolumn {8}{c}{all durations in hours, $\lambda_O$ in per hour}
	\end{tabular}
	\vspace{-3mm}
\end{table}

On average, events in the Eastern interconnection are larger than in the West and ERCOT. It can be explained by the fact that the largest transmission events were caused by hurricanes, and all of these events occurred in the East. 
For all interconnections, the mean and median outage process durations $D_O$ are similar, and very short compared to event durations $D_E$.
The mean outage rate in the West is much higher due to several events (wildfires and a lightning storm) for which all outages started almost simultaneously. 
This extremely short outage duration $D_O$ results in huge outage rates (see (\ref{lambdaOhat})).

The restoration usually starts very quickly after the event starts as the time to first restore $D_{r1}$ indicates. In ERCOT the average time to a first restore, 1 hour 17 minutes, is statistically significantly larger than in the East and in the West, where restoration typically starts within one hour. 
Overall, the time to first restore is negligible compared to event duration; this makes the event duration $D_E$ and the restore process duration $D_n$ effectively equal.
In contrast, the time between the $(n-1)$th and $n$th restores, $D_n-D_{n-1}$, is sizeable and often comprises a substantial share (41\% on average) of  $D_n$. 
This observation again underscores the impact of the last few restores to the event and restore durations. 

The geometric mean of the positive restore times, $D_{\rm GM}$, is a simple and stable metric. $D_{\rm GM}$ is also an approximate estimate for the time to one half of restores for the events with log-normal restore times. 
The largest difference between these metrics observed for the ERCOT events can be attributed to the poorer log-normal fit  for the ERCOT events. 
On average, $D_{\rm GM}$ is  12\%  of the entire  restore process duration $D_n$.

It is interesting to compare in Table \ref{typical} the sample quantile  restore time $D_{95\%}$ with the lognormal and exponential quantiles $D_{95\%}^{\rm ln}$ and $D_{95\%}^{\rm exp}$. $D_{95\%}^{\rm ln}$ often overestimates $D_{95\%}$   due to the heavy tail of the lognormal distribution,
whereas $D_{95\%}^{\rm exp}$ often  underestimates $D_{95\%}$ due to the light tail of the exponential distribution.

The parameters $\mu$ and $\sigma$ for fitted log-normal distributions and $\tau$ for fitted exponential are consistent in each interconnection and across interconnections. 
Table~\ref{CIsizetable} shows that $\mu$ increases and $\sigma$ decreases with event size $n$.

Only 23 of the 352 resilience events in  the  dataset  are not weather-related. 
These 23 events vary in size from  10 to 26 outages. 
Except for $D_{r1}$,
the medians of the duration metrics in Table~\ref{weathertype} are statistically significantly higher\footnote{confirmed with a nonparametric one-way ANOVA test for medians \cite{Hollander99}} for weather-related events than for non weather-related events. 
Table~\ref{weathertype} also shows for each weather type the median metrics  for the 95 weather-related events with  at least 18 outages. 
There are some
 statistically significant differences$^7$ among the extreme weather types:  
the medians of $D_n$ and $D_{95\%}$ for hurricanes are  greater than for other weather types, and $D_{\rm GM}$ and $\mu$ for hurricanes and tornadoes are  greater than for other weather types. 
The mean of the times to first restore $D_{r1}$ are similar for all weather types except tornadoes; the mean
$D_{r1}$ for tornadoes is 1.7 hours, which is at least double the mean $D_{r1}$ for the other  weather types.

\begin{table}[ht]
	\caption{Median value of metrics by type of weather}
	\label{weathertype}
	\centering
	\setlength{\tabcolsep}{0.5em}
	\begin{tabular}{ rccccccccc}
\hspace{-4mm}Type (\# cases)\hspace{-2mm}~&  $n$&$D_O$ &$D_{r1}$&$D_{n}$ &$D_{\rm GM}$&$D_{95\%}$&$\mu$&$\sigma$ \\
		\hline
fire \phantom{3}(4) &21	&1.51	&0.33	&33.4	&2.63		&30.8	&0.96	&1.89\\
hurricane (17)	&55	&6.53	&0.58	&257	&20.4		&109	&3.02 &1.50\\
~\!\!\!\!\!wind,thunder (36)	&25	&4.04	&0.44	&122	&6.75		&82.3	&1.90	&1.44\\
tornado (15)	&24.5	&5.04	&0.96	&174	&12.7		&93.4	&2.54	&1.47\\
winter (23)		&32	&4.37	&0.60	&49.5	&4.73		&41.5	&1.55	&1.32\\
all weather\,(329)	&14	&2.80	&0.52	&73.4	&5.76		&67.7	&1.75	&1.56\\
~\!\!\!\!non-weather (23)	&11	&1.00	&0.65	&19.1	&1.10		&19.1	&0.09	&1.58\\
					        		\hline			\\[-3mm]
					        	&&	\multicolumn {5}{c}{all durations in hours}
	\end{tabular}
	\vspace{-3mm}
\end{table}

Our analysis confirms a well-known fact that a type of extreme weather can be more typical and impactful for one interconnection than another. Among the 11 named hurricanes that caused 17 transmission outage events shown in Table \ref{weathertype} (the largest, longest and most impactful events in the data set) all except one hit the Eastern Interconnection; the exception was the hurricane Harvey (ERCOT, August 2017). Wildfires causing large transmission events usually occur in the West. These examples demonstrate a possible reason in metric variability across the system and, more importantly, the impractically of using duration metrics to compare resilience of transmission system in different interconnections. These metrics should be used to track differences in resilience and restoration for the same grid (changes in time, between different types of events etc.).

\section{Fit of Poisson process models to utility data}
\label{fit}

This section discusses the fit of the Poisson models to the observed utility data by a goodness of fit test, which allows for analysis of each of the 352 events, and by probability plots for the combined normalized data, which also show where the fit deviates. For the goodness of fit tests,
there is some arbitrariness in the threshold amount of deviation corresponding to the significance level, as well as some dependence on the event size $n$, but they do give an indication of fit.

\subsection{Outage process fit with uniform distribution}
The Poisson process model with constant outage rate implies that for each event the $n\!-\!2$ outage times $o_k$, $k=2,3, ..., n\!-\!1$  should be independent samples from a uniform distribution on the interval $(o_1,o_n)$.
We  evaluated  the fit of these outage times for each event to the uniform distribution as shown in Table~\ref{percentevents}.
Satisfying the test means that the ideal model is not rejected at the significance level $0.05$.
Table~\ref{percentevents} shows that a majority of events satisfy the model.
 \begin{table}[ht]
	\caption{\footnotesize Percent of events satisfying outage and restore models }
	\label{percentevents}
	\centering
	\begin{tabular}{rcccc}
\multicolumn{1}{c}{test}	& \multicolumn{4}{c}{interconnection}\\
		\multicolumn{1}{c}{(satisfies if $p\ge0.05$)}& ~~~all~~~&eastern& western&{\textsc{ercot}}\\
		\hline
		\multicolumn{5}{l}{\textsc{percent of events satisfying uniform outages}{\vrule height 8pt depth 0pt width 0pt}}\\
	       Kolmogorov-Smirnoff &69&70&72&50	\\
	       Cramer-vonMises     &72&73&71&56\\
	       Anderson-Darling    &63&63&66&50\\[2pt]
	       \multicolumn{5}{l}{\textsc{percent of events satisfying lognormal restores}}\\
	        Kolmogorov-Smirnoff &63&63&66&44\\
	       Cramer-vonMises &60&61&64&38\\
	       Anderson-Darling &59&59&64&38\\[2pt]
	       \multicolumn{5}{l}{\textsc{percent of events satisfying exponential restores}}\\
	       Kolmogorov-Smirnoff &35&33&42&25\\
	       Cramer-vonMises &35&33&45&25\\
	       Anderson-Darling &32&31&40&13
			\end{tabular}
\end{table}

\begin{figure}[htbp]
	\centering
	\includegraphics[width=0.75\columnwidth]{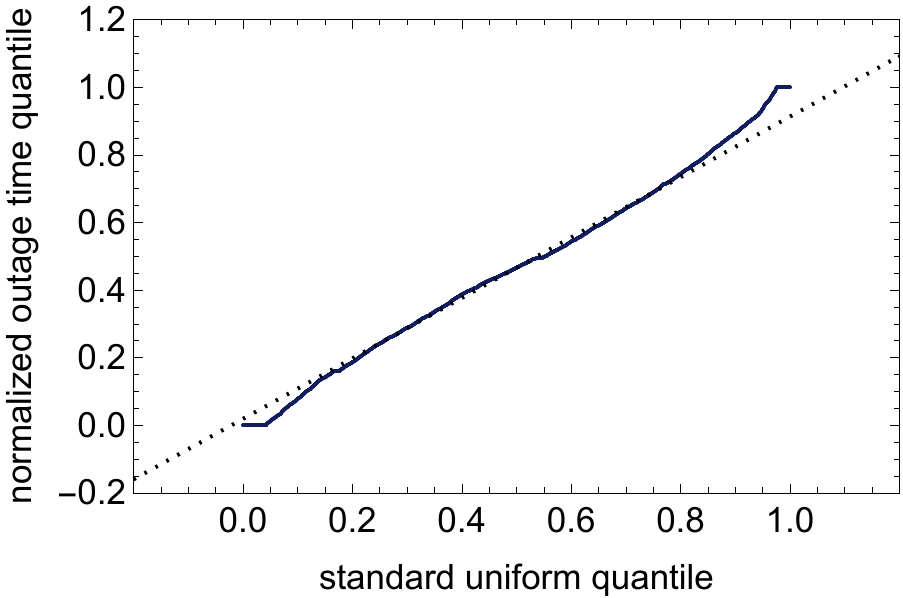}
	\caption{Fit of normalized outage data to standard uniform distribution on QQ plot.}
	\label{QQoutageEastern}
\end{figure}

The normalized  outage times $(o_k-o_1)/(o_n-o_1)$, $k=2,3,...,n\!-\!1$ should be independent samples from the standard uniform distribution on the interval $(0,1)$.
The fit of the normalized outage times for all of the events to the standard uniform distribution is shown by the QQ plot in Fig.~\ref{QQoutageEastern}. 
The fit in Fig.~\ref{QQoutageEastern} is quite close over the middle range, and 
the main deviations occur at the ends of the distribution and correspond to simultaneous multiple outages recorded at the beginning or end of the outage process\footnote{While it is plausible that some outage processes start or end with outages occurring in the same minute, it is not clear that the records accurately reflect the outage timing in all these cases.}.

The fits of this subsection indicate that the Poisson model with uniform rate is a typical case (a majority of all events) usefully approximating the outage process.

\begin{figure}[htbp]
	\centering
	\includegraphics[width=0.75\columnwidth]{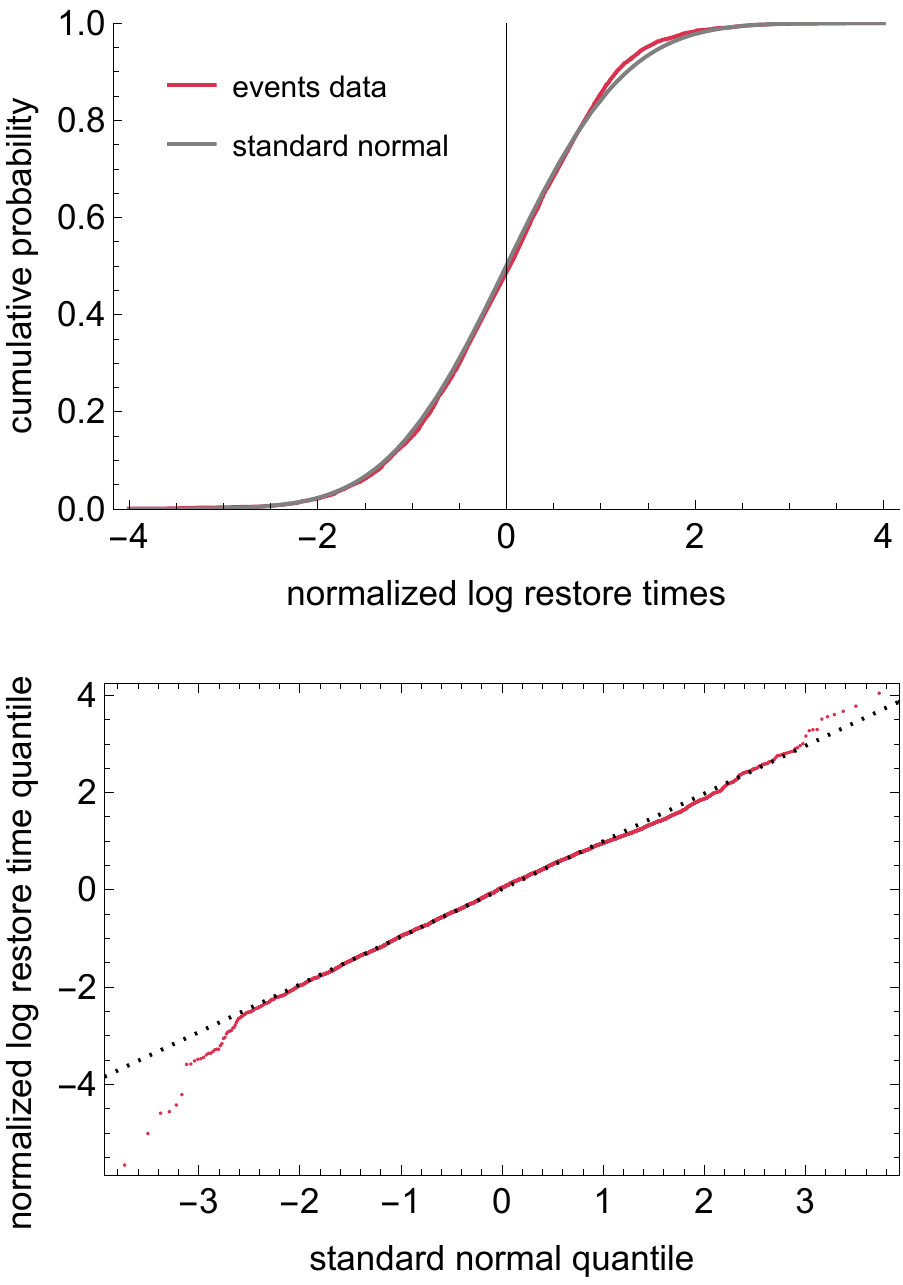}
	\caption{Fit of normalized log restore data to standard normal distribution. Above compares CDFs; below is QQ plot.}
	\label{CDFrestoreEastern}
\end{figure}

\begin{figure}[htbp]
	\centering
	\includegraphics[width=0.75\columnwidth]{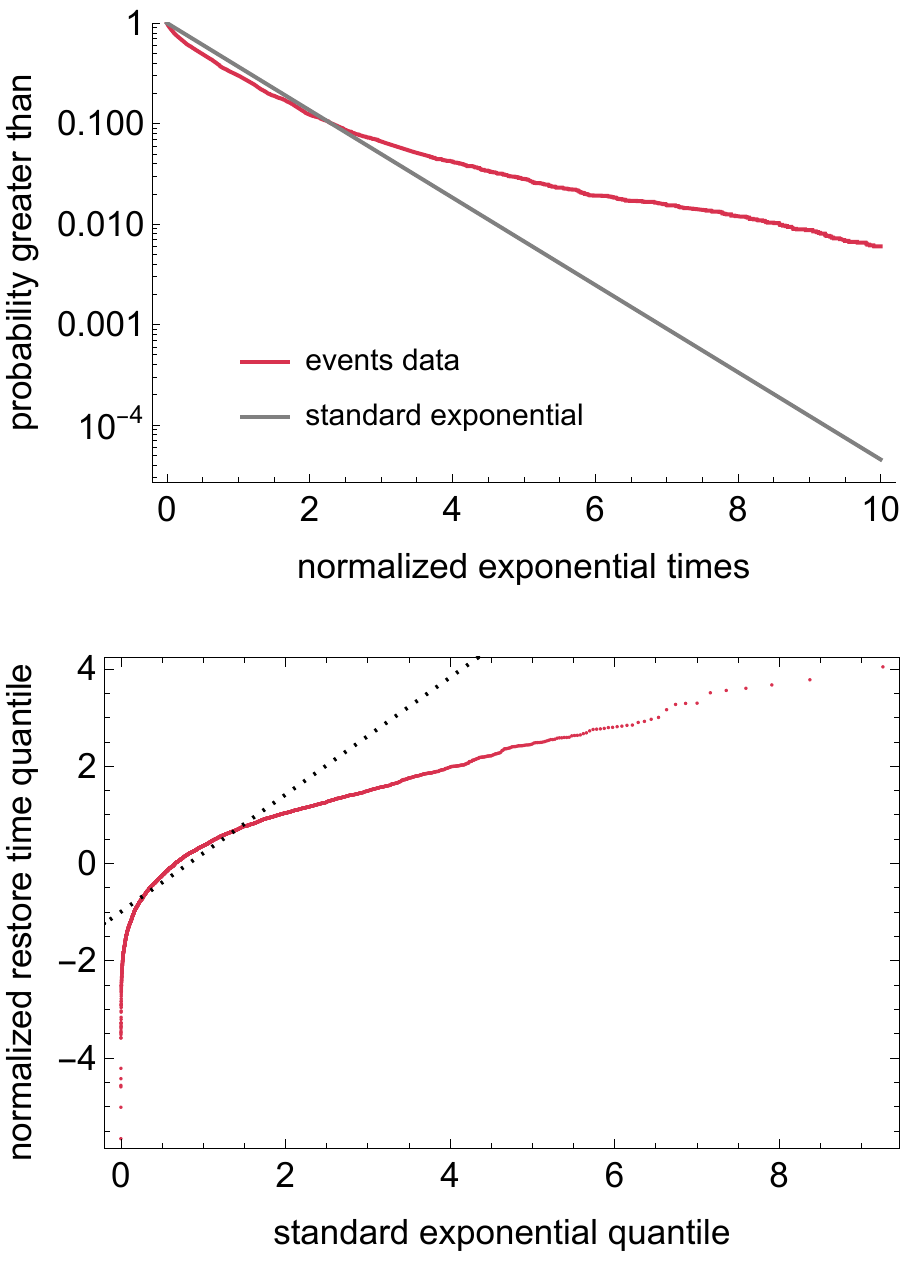}
	\caption{Fit of normalized restore data to standard exponential distribution. Above compares log survival functions; below is QQ plot.}
	\label{CDFexprestoreEastern}
\end{figure}

\subsection{Restore  process fit with lognormal distribution}
As explained in section \ref{Poissonlognormal}, the Poisson process model with lognormal rate for the restores  implies that for each event the restore times 
$r_{z+1}\!-\!r_1,r_{z+2}\!-\!r_1,...,r_n\!-\!r_1$  should be independent samples from a lognormal distribution. 
We  evaluated  the fit of these restore times for each event to the lognormal distribution with parameters $\mu,\sigma$ estimated
using (\ref{muhat}), (\ref{sigmahat})
at the significance level  $0.05$ as shown in Table~\ref{percentevents}.
Table~\ref{percentevents} shows that a majority of all events satisfy the model, and this also holds for the East and West interconnections.

For each event, 
 the normalized  restore times $(\ln(r_k\!-\!r_1)-\mu)/\sigma$, $k=z+1,z+2,...,n$ should be independent samples from the standard normal distribution $N(0,1)$.
The fit of the normalized restore times for all events to the standard normal distribution is shown by the CDF and QQ plots in Fig. \ref{CDFrestoreEastern}, which show a reasonably good  fit with some modest deviations.

The fits described in this subsection indicate that the Poisson process model with lognormal rate is a typical case usefully approximating the restore process. The typical lognormal case has a  heavy tail that can describe some extremely delayed final restores.

\subsection{Restore  process fit with exponential distribution}
As explained in section \ref{Poissonexp}, the Poisson process model with exponential rate for the restores  implies that for each event the  restore times
$r_{z+1}-r_1,r_{z+2}-r_1,...,r_n-r_1$  should be independent samples from an exponential distribution with time constant $\tau$.
We evaluate  the fit of the restore times for each event to the exponential distribution with time constants $\tau$ estimated using (\ref{tauhat}) 
as shown in Table~\ref{percentevents}.
Table~\ref{percentevents} shows that a minority of events satisfy the model.

For each event,
 the normalized  restore times $\tau^{-1}(r_k\!-\!r_1)$, $k=z+1,z+2,...,n$ should be independent samples from the standard exponential distribution with time constant 1.
The fit of the normalized restore times for all  events to the standard exponential distribution is shown by the survival function and QQ plots in Fig.~\ref{CDFexprestoreEastern}. 
There is clear discrepancy between the exponential model and the data for the initial portion and tail of the distribution.
The tail in the data is much heavier than exponential, and 
this discrepancy in the tail is particularly significant for our purpose here of estimating restore durations.

The fits described in this subsection indicate that the Poisson process model with exponential rate only fits a minority of the events and is a noticeably poorer approximation of the typical restore process than the  model with lognormal rate.

\section{Stochastic variability of restore metrics}
\label{vary}

The restore duration metrics vary due to variation of the restore processes between events (and this of course is what we want to quantify) but also due to the inherent statistical variability of the metric used (which we want to minimize by selecting a better metric). 
The statistical variability makes the metric vary between events, even if the events have the same characteristics, because of random variations in the progress of the restores. 

We  assess the inherent statistical variability of the metrics by assuming the lognormal Poisson model for average values of $\mu$  and $\sigma$, which vary as  functions of $n$, and are estimated using (\ref{muhat}) and (\ref{sigmahat}). 
In this section we assume that $z=1$.

\subsection{Variability of $D_{GM}$, $D_{x\%}^{\ln}$, $\tau$}
Since $z=1$ is assumed, $\mu$ and $\sigma$ are estimated with $n\!-\!1$ samples.
The sample mean $\mu$ of $n\!-\!1$ samples from a normal distribution with mean $\mu$ and standard deviation $\sigma$ has normal distribution $N(\mu,\sigma/\sqrt{n\!-\!1})$.
Therefore $\mu$ 
has two-sided $100(1-c)\%$ confidence interval with end points $\mu\pm\sigma z_c/\sqrt {n\!-\!1}$, where $z_c= \Phi^{-1}(1-c/2)$ and $\Phi$ is the CDF of the standard normal distribution. 
It follows that the geometric mean $D_{\rm GM}$ of $n\!-\!1$ samples from a lognormal distribution with parameters $\mu$ and  $\sigma$ has two-sided $100(1-c)\%$ confidence interval with endpoints $\exp[\mu\pm\sigma z_c/\sqrt {n\!-\!1}\,]$, or
\begin{align}
 \big \{ e^{\mu}\div\exp(\sigma z_c/\sqrt {n\!-\!1}\,),\,e^{\mu}\times\exp(\sigma z_c/\sqrt {n\!-\!1}\,)\big\}
    \label{CIDGM}
\end{align}
We measure the size of the $D_{\rm GM}$ confidence interval (\ref{CIDGM}) by the multiplicative factor $\exp(\sigma z_c/\sqrt{n\!-\!1}\,)$, which we call 
  the ``multiplicative half-width" of the confidence interval.
More generally, we define the  size of a confidence interval
with endpoints $c_1, c_2$ as
\begin{align}
\mbox{multiplicative half-width of }\{c_1, c_2\}=\sqrt{c_2/c_1}
\label{multhalfwidth}
\end{align}

Now we obtain the size of the confidence interval for $D_{x\%}^{\rm ln}$.
From (\ref{Dxpercent}), taking $z=1$,
\begin{align}~\hspace{-4mm}
\ln D_{x\%}^{\rm ln}&=\mu+ \phi_{x,n}\sigma,\,
\mbox{where }\phi_{x,n}=\Phi^{-1}\Big[\frac{nx/100-1}{n-1}\Big]\label{lnDxln}
\end{align}

The sample standard deviation 
$\sigma$ has distribution $(\sigma/\sqrt{n\!-\!2})\chi_{n\!-\!2}$ where $\chi_{n\!-\!2}$ is the chi distribution with $n\!-\!2$ degrees of freedom\footnote{the definition of $\sigma$ uses $\mu$, so that the number of degrees of freedom is one fewer than the number of samples $n\!-\!1$}.

Using (\ref{lnDxln}) and the independence of $\mu$ and $\sigma$,
the probability density function of $\ln D_{x\%}^{\rm ln}$ is the convolution
\begin{align}
    f_{\ln\!D_{x\%}^{\rm ln}}=f_{N(\mu,\sigma/\sqrt{n\!-\!1})}*f_{(\phi_{x,n}\sigma/\sqrt{n\!-\!2})\chi_{n\!-\!2}}
    \label{flnDxln}
\end{align}
and the CDF of $\ln D_{x\%}^{\rm ln}$ is
\begin{align}
    F_{\ln\!D_{x\%}^{\rm ln}}=
    f_{N(\mu,\sigma/\sqrt{n\!-\!1})}*F_{(\phi_{x,n}\sigma/\sqrt{n\!-\!2})\chi_{n\!-\!2}}
    \label{FlnDxln}
\end{align}
We use (\ref{FlnDxln}), numerically integrating to evaluate the
convolution, to find the   $100(1-c)\%$ confidence interval for $\ln D_{x\%}^{\rm ln}$ as  $\{F_{\ln\!D_{x\%}^{\rm ln}}^{-1}(c/2),F_{\ln\!D_{x\%}^{\rm ln}}^{-1}(1-c/2)\}$,
then use (\ref{multhalfwidth}) to find the  multiplicative half-width of the confidence interval for  $D_{x\%}^{\rm ln}$.

Equation (\ref{tauhat}) shows that the exponential time constant $\tau$ is also the arithmetic mean of the nonzero restore times.  In this section these $n-1$ restore times are assumed to be sampled from a lognormal distribution.
Using Cox's approximate method \cite{ZhouSIM97}, the multiplicative half-width of the confidence interval of $\tau$ is
\begin{align}
    \exp\big(z_c\sigma\sqrt {1/(n-1)+\sigma^2/(2n-4)}\,\big)
\end{align}

\subsection{Variability of $D_k$ and $D_{x\%}$}
\looseness=-1
Since the restore times $r_1,r_2,...,r_n$ are sorted in increasing order, $D_k=r_k-r_1$ corresponds to the $k$th largest restore time and, assuming that $z=1$ and $k\ge2$, $D_k$ is the $(k\!-\!1)$th order statistic of 
the $n\!-\!1$ lognormally distributed restore times $r_2-r_1,...,r_n-r_1$.
We evaluate in Mathematica the inverse CDF $F_{D_k}^{-1}$ of the $(k\!-\!1)$th order statistic of $n\!-\!1$ samples of the lognormal distribution with parameters $\mu$ and $\sigma$.
Then we find the $100(1\!-\!c)\%$ confidence interval for $D_k$
$\{F_{D_k}^{-1}(c/2),F_{D_k}^{-1}(1\!-\!c/2)\}$ and its multiplicative half-width  from (\ref{multhalfwidth}).

To evaluate the variability of $D_{x\%}$, we approximate its inverse CDF with the linear interpolation 
\begin{align}
F_{D_{x\%}}^{-1}(p)=(1- (u-\lfloor u\rfloor))F_{D_{\lfloor u\rfloor}}^{-1}(p)+(u-\lfloor u\rfloor)F_{D_{\lceil u\rceil}}^{-1}(p)
\notag
\end{align}
where $u$ is given by (\ref{u}).
We  then  obtain the   $100(1-c)\%$ confidence interval
$\{F_{D_{x\%}}^{-1}(c/2),F_{D_{x\%}}^{-1}(1-c/2)\}$ and use (\ref{multhalfwidth}) to obtain its multiplicative half-width.

\subsection{Results for variability of metrics}
The size of the 90\% confidence interval, measured by the multiplicative half-width (\ref{multhalfwidth}), indicates the inherent statistical variability of the metrics. For example, a multiplicative half-width of 2 indicates that the interval spans from  half to double of a point inside the interval.
Table~\ref{CIsizetable} shows results for metric variability, and
there are some overall trends:
All the metrics become much more variable as the event size $n$ decreases.
Metrics estimating a larger fraction of the entire restore duration are much more variable
(consider the sequence $D_{\rm GM}=D_{50\%}^{\rm ln}$, $D_{90\%}^{\rm ln}$, $D_{95\%}^{\rm ln}$ or $D_{50\%}$, $D_{90\%}$, $D_{95\%}$, $D_{100\%}=D_n$).
The quantile metrics ($D_{50\%}$, $D_{90\%}$, $D_{95\%}$) are always more variable than corresponding metrics related to lognormal restore ($D_{\rm GM}$, $D_{90\%}^{\rm ln}$, $D_{95\%}^{\rm ln}$), but the increase in variability is modest or small for $n\ge$50.

\begin{figure}[htb]
	\centering
	\includegraphics[width=\columnwidth]{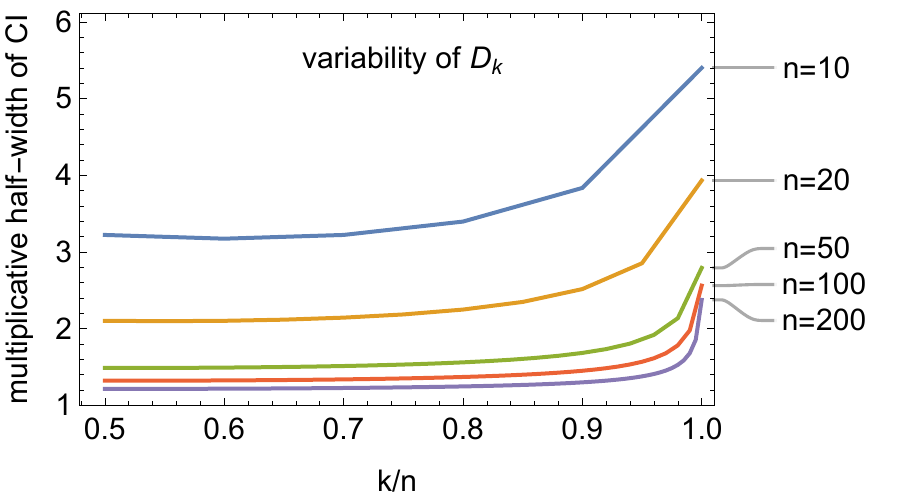}
	\caption{Size of 90\% confidence interval for $k$th order statistic $D_k$ as the fraction $k/n$ varies. $n$ is number of restores.
	Confidence interval size is multiplicative half-width. Lognormal restore is assumed with parameters $\mu$ and $\sigma$ from Table~\ref{CIsizetable}.}
	\label{CIsizeplot}
\end{figure}

Metric variability is worst and unacceptably large for
$D_n$, which always has a confidence interval size of  more than a factor of 2. The high variability of the  last restore $r_n$ and $D_n$ is expected due to the heavy tail of the lognormal distribution.
Fig.~\ref{CIsizeplot} shows that the variability of $D_k$ is sharply reduced for $k/n=0.95$, at least for larger $n$, and further reduced for $k/n=0.90$.
This motivates avoiding $D_n$ and considering the use of $D_{90\%}^{\rm ln}$, $D_{95\%}^{\rm ln}$, $D_{90\%}$, $D_{95\%}$, which have confidence intervals with size less than a factor of 2 for $n\ge50$ and which perform more continuously by interpolating the $D_k$ metrics. The arithmetic mean $\tau$ is highly variable for smaller values of $n$; it  has a confidence interval size of more than a factor of 2 for $n<30$.

The pervasive problem of duration metric variability is best mitigated by $D_{\rm GM}$, which has a confidence interval size of less than a factor of 2 for $n\ge17$.

\begin{table*}[tbp]
\caption{multiplicative half width of 90\% confidence interval}
\label{CIsizetable}
\centering
\setlength{\tabcolsep}{0.4em}
\begin{tabular}{ccc|ccccccccc}
 $n$&$\mu$&$\sigma$ & $D_{\rm GM}$ &  $D_{50\%}$ &$\tau$& $D^{\rm ln}_{90\%}$&$D_{90\%}$& $D^{\rm ln}_{95\%}$&$D_{95\%}$&$D_{n-1}$ & $D_{n}$ \!\!\!
 	{\vrule height 12pt depth 5pt width 0pt}\\
 \hline
 	{\vrule height 9pt depth 0pt width -3pt}
10 & 1.18 & 1.72 & 2.57 & 3.19 & 4.68 & 3.56 & 5.09 & 4.29 & 5.40 & 3.83 & 5.40 \\
 20 & 1.60 & 1.58 & 1.82 & 2.10 & 2.49 & 2.24 & 2.76 & 2.51 & 3.69 & 2.85 & 3.93 \\
 50 & 2.20 & 1.35 & 1.37 & 1.49 & 1.56 & 1.54 & 1.72 & 1.63 & 1.96 & 2.14 & 2.79 \\
 100 & 2.52 & 1.35 & 1.25 & 1.32 & 1.36 & 1.35 & 1.46 & 1.41 & 1.60 & 1.98 & 2.56 \\
 200 & 3.15 & 1.33 & 1.17 & 1.21 & 1.24 & 1.23 & 1.30 & 1.27 & 1.39 & 1.85 & 2.37 \\
\end{tabular}
\end{table*}

This section assesses metric variability assuming the lognormal model of restores. This is a good assumption for a majority of cases, and can be regarded as a stringent assumption for the remaining minority of cases due to the heavy tail of the lognormal distribution.

\section{Conclusions}

\looseness=-1
We use extensive North American transmission system  data to analyze the statistical variability and  interpretations of
a variety of metrics for the duration of processes in resilience events. 
Some metrics, such as the outage duration $D_O$, outage rate $\lambda_O$, and the time delay  before the first restore $D_{r1}$, are useful. 
Other duration metrics can suffer from excessive statistical variability, in which their estimated values are contained in confidence intervals that are so large that the estimated values of the metric are not representative.
 This variability is quantified using new Poisson models for outage and restore processes. The variability is worse for small events.
 
The apparently straightforward metrics of restore process duration $D_n$ and the event duration $D_E$ are extremely statistically variable and do not adequately describe the restore process, so we recommend new duration metrics $D_{\rm GM}$ and $D_{95\%}$ (or $D_{90\%}$) with better performance. 
In particular,
the geometric mean of restore times $D_{\rm GM}$ has the least statistical variability, summarizes all of the restore process, and approximates a time at which half the restores are completed.
The quantile-based metric $D_{95\%}$ indicates the time at which restoration is 95\% complete, but has
 greater variability than $D_{\rm GM}$.
$D_{95\%}$ uses interpolation to vary more continuously as the data changes. Table~\ref{summary} summarizes the metrics and their recommendations, and  Tables~\ref{typical} and \ref{weathertype} give typical values for the metrics for three interconnections and different weather conditions.

Since our paper is driven by North American bulk electric transmission system outage data, strictly speaking the results describe aspects of resilience only in North American transmission grids. However, since similar transmission outage data is routinely collected worldwide, the methods of the paper are readily applicable to other transmission systems to test or confirm the models and conclusions of the paper.

We introduce novel Poisson process models for the outage and restore processes in resilience events. These new stochastic models describe how resilience events progress in North American transmission systems, and are verified with extensive utility data to be good approximations for the majority of cases.
The outages occur uniformly over a short time interval, whereas the restores occur at a lognormal rate that slows to produce the long delays often observed for the last few restores. 
 The lognormal model for the restores is a noticeably better fit than an exponential model for the restores.
 We give typical values of the model parameters for three interconnections and for different weather conditions to make the new models more specific and useful to other researchers.
 
The Poisson process models describe probabilistic outages and restores occurring according to specified rates. Averaging the Poisson process models produces formulas for smooth, deterministic curves that approximate typical outage and restore processes. These deterministic averaged models are of considerable interest for future work describing how resilience events
progress in transmission systems. For example, one can derive formulas for the area, duration, and nadir metrics of mean performance curves in terms of the Poisson process parameters \cite{DobsonArXiv23}. The formulas for area under the mean performance curve are simple and intuitive, and sometimes also apply to the area under empirical performance curves that are obtained from observed data.

\bibliographystyle{IEEEtran}

   \begin{IEEEbiographynophoto}{Ian Dobson}
   (IEEE Fellow) received the B.A. in mathematics from Cambridge University and the Ph.D. in electrical engineering from Cornell University.
He previously worked for British industry and  the University of Wisconsin-Madison, and is currently Sandbulte Professor of Electrical  Engineering at Iowa State University. His interests are power system resilience, blackout risk, cascading failure, complex systems, and nonlinear dynamics.
    \end{IEEEbiographynophoto}

 \begin{IEEEbiographynophoto}{Svetlana Ekisheva} (IEEE Senior Member) is a principal data science advisor at the North American Electric Reliability Corporation (NERC). She is responsible for developing statistical models and performing statistical analysis of the databases supported by NERC. Her expertise also includes 15 years of teaching and conducting research at universities in Russia and the U.S. Svetlana earned her Ph.D. in Probability and Statistics from St.-Petersburg State University, Russia. She is a member of American Statistical Association. Her current research interests are in risk assessment, reliability and resilience analysis of the Bulk Power System. 
   \end{IEEEbiographynophoto}

\end{document}